\documentclass[aps,prl,showpacs,twocolumn,floatfix,superscriptaddress]{revtex4-1}

\usepackage{amsmath,amssymb}
\usepackage{graphicx}
\usepackage{dcolumn}
\usepackage{bm}
\usepackage[usenames]{color}
\usepackage[breaklinks=true,colorlinks=true,linkcolor=blue,urlcolor=blue,citecolor=blue]{hyperref}
\usepackage{ulem}
\usepackage{amsmath}
\usepackage{ulem}

\usepackage{amsfonts}%
\usepackage{amsthm}%
\usepackage{mathrsfs}%
\usepackage{gensymb}%

\begin{document}

\date{\today}

\title{Electroosmotic Flow in Different Phosphorus Nanochannels}
%% Active Semi-flexible Polymer: Structure, Dynamics and Bifurcation}
%\thanks{Footnote to title of article.}

\author{S.M.Kazem Manzoorolajdad}
\email{kazem.manzoor@email.kntu.ac.ir}
\affiliation{Department of Physics, K.N. Toosi University of Technology, Tehran, 15875-4416, Iran}
\affiliation{School of Nano sciences, Institute for Research in Fundamental Sciences (IPM), 19395-5531, Tehran, Iran.}

\author{Hossein Hamzehpour}
\email{hamzehpour@kntu.ac.ir}
\affiliation{Department of Physics, K.N. Toosi University of Technology, Tehran, 15875-4416, Iran}
\affiliation{School of Nano sciences, Institute for Research in Fundamental Sciences (IPM), 19395-5531, Tehran, Iran.}

\author{Jalal Sarabadani}
\email{jalal@ipm.ir}
\affiliation{School of Nano sciences, Institute for Research in Fundamental Sciences (IPM), 19395-5531, Tehran, Iran.}

%\date{\today}% It is always \today, today,
             %  but any date may be explicitly specified

\begin{abstract}
The electrokinetic transport in a neutral system consists of an aqueous NaCl solution confined in a nanochannel with two similar parallel phosphorene walls, and is investigated for different black, blue, red, and green phosphorene allotropes in the presence of an external electric field in the directions $x$ (parallel to the walls roughness axis) and $y$ (perpendicular to the walls roughness axis). 
The results show that irrespective of the electric field direction, the thickness of the Stern layer increases with the increase in the magnitude of the negative electric surface charge density (ESCD) on the nanochannel walls, and it also increases with the increase in the roughness ratio for different allotropes. Moreover, three different regimes of Debye--H\"{u}ckel (DH), intermediate, and flow reversal appear as the absolute value of the negative ESCD on the walls grows. With the increase in the absolute value of the negative ESCD, in the DH regime, the flow velocity grows, then in the intermediate regime, it decreases, and finally, at sufficiently high ESCD, the flow reversal occurs. When the external electric field is applied in the $y$ direction, the dynamics of the system are slower than that of the $x$ direction; therefore, the flow reversal occurs at the smaller absolute values of the negative ESCD.
\end{abstract}

\maketitle

%%%%%%%%%%%%%%%%%%%%%%%%%%%%%%%%%%%%%%%%%%%%%%%%%%%%%%%%%%%%%%%%%%%%%%%%%%%%%%%%%%%%%%%%%%%%%%%%%%%%%%
%%%%%%%%%%%%%%%%%%%%%%%%%%%%%%%%%%%%%%%%%%%%%%%%%%%%%%%%%%%%%%%%%%%%%%%%%%%%%%%%%%%%%%%%%%%%%%%%%%%%%%
%%%%%%%%%%%%%%%%%%%%%%%%%%%%%%%%%%%%%%%%%%%%%%%%%%%%%%%%%%%%%%%%%%%%%%%%%%%%%%%%%%%%%%%%%%%%%%%%%%%%%%

\section{INTRODUCTION} \label{introduction}

Electrokinetics plays an important role in the dynamics of liquid, ion, or bio--particles in micro/nano scale systems. It has broad applications ranging from desalination to drug delivery, micro/nano pumping, geosciences, and energy storage \cite{cohen2012water,angelova2008dynamic,ramsey1997generating,hamzehpour2014electro,merlet2014electric,%
zhang2022diffusio,hardt2020electric,faraji2020electrokinetic,chowdhury2022electrophoresis}. Various factors may drive the fluid flow inside nanochannels, e.g., pressure, surface tension gradient, electric field, temperature, and concentration gradients \cite{squires2005microfluidics,schoch2008transport}. Electroosmotic flow (EOF) offers a fast and effective way to control the flow in highly confined environments compared to other driving forces. The EOF is the motion of ionic liquids drived by applying an external electric field on the net charge inside the electric double layer (EDL) in the system. When an ionic solution is placed in contact with quenched charges on a dielectric surface, the surface is rapidly charged due to the protonation/deprotonation reactions with the ionic solution. This leads to a net-charged EDL, which is formed by counterions  \cite{karniadakis2006microflows}. The excess of dissolved ions in the diffuse layer of the EDL flow with or against the applied electric field and drag the surrounding water molecules. The EDL thickness for most practical applications ranges from 1~\!nm for high conductivity solutions to 100~\!nm for deionized water \cite{dutta2002electroosmotic}. 

In nanofluidic systems, the classical continuum assumptions and the mean-field Poisson–Boltzmann (PB) theory may not adequately describe the flow field and the ion distribution. For instance, the classical PB formalism that solely considers electrostatic interactions between charged species could not provide an adequate explanation of high surface charges and ion valency. In addition, fluctuations of ions and water densities near the surface and discrete effects due to finite molecular sizes, which are not considered in the original PB theory, are non-negligible in narrow fluidic environments \cite{silalahi2010comparing,tresset2008generalized,fleck2005counterion,boroudjerdi2005statics}. The flow reversal and some anomalous transport phenomena caused by charge inversion and ion-specific effects are not taken into account by the continuum models~\!\cite{qiao2004charge,huang2007ion}. In some works, people make modifications to classical theories to obtain accurate predictions \cite{celebi2018molecular,celebi2019molecular}. However, molecular dynamics (MD) simulations give more accurate results by taking into account the details at the molecular level. For example, Holt {\it et al.} showed that the flow velocity inside carbon nanotubes is more than three orders of magnitude larger than predicted by continuum hydrodynamic models with stick boundary conditions~\!\cite{holt2006fast}.

The electroosmosis and electric double layer characteristics are significantly affected by the induced surface charge, nanochannel wall surface parameters such as geometry and wettability, types of ions, ion concentration, and surface chemistry. The effects of surface charge on the surface wettability and wetting kinetics have been studied by Puah {\it et al.}~\!\cite{puah2010influence}. An increase in the surface charge due to the effect of a stronger wall force field increases the number of molecules at the interface, which leads to an increase in the surface wettability. Qiao \textit{et al.}~\! \cite{qiao2007effects} have investigated the EOF in the nanochannel in the presence of surface roughness at the molecular level. They showed that the surface roughness strongly changes the ion distribution and electroosmotic velocity when the roughness height is comparable to the thickness of the  EDL. Kim and Darve~\!\cite{kim2006molecular} have studied the EOF in rough nanochannels with artificial rectangular surface roughness and showed that with increasing the roughness, the flow velocity and zeta potential decrease. Another model of a rough surface with a fractal structure was proposed by Lu \textit{ et al.}~\!\cite{lu2017electroosmotic}, wherein it has been shown that for a definite roughness height, the flow velocity is strongly affected by the surface fractal dimensions and decreases monotonously with the increasing of the roughness height.

Almost all previous studies on electroosmosis in nanochannels with rough surface walls have been performed with artificial rough surfaces. There are some rough two-dimensional materials in nature that can be used as boundary walls of a nanochannel. One important natural two-dimensional rough material is phosphorene. To the best of our knowledge, there has not been any study on electrosmosis in a nanochannel with phosphorene walls. In this paper, the effects of the surface charge density, direction of the external electric field, and morphology of four common black, blue, red, and green phosphorene allotropes as nanochannel walls on the EOF velocity and Stern layer are studied. Our results show that the surface geometry and roughness of these allotropes are able to significantly affect the fluid flow behavior. The rest of this paper is organized as follows. 
In the next section, the geometrical and physical properties of phosphorene allotropes are described. Then the details of the molecular dynamics simulations are explained. After that, the results are presented and discussed. The last section is devoted to a summary.

%%%%%%%%%%%%%%%%%%%%%%%%%%%%%%%%%%%%%%%%%%%%%%%%%%%%%%%%%%%%%%%%%%%%%%%%%%%%%%%%%%%%%%%%%%%%%%%%%%%%%%
%%%%%%%%%%%%%%%%%%%%%%%%%%%%%%%%%%%%%%%%%%%%%%%%%%%%%%%%%%%%%%%%%%%%%%%%%%%%%%%%%%%%%%%%%%%%%%%%%%%%%%

\section{GEOMETRY OF NANOCHANNELS WALL} \label{geometry}

\begin{figure*}[t]
	\begin{minipage}{1.0\textwidth}
    \begin{center}
        \includegraphics[width=0.58\textwidth]{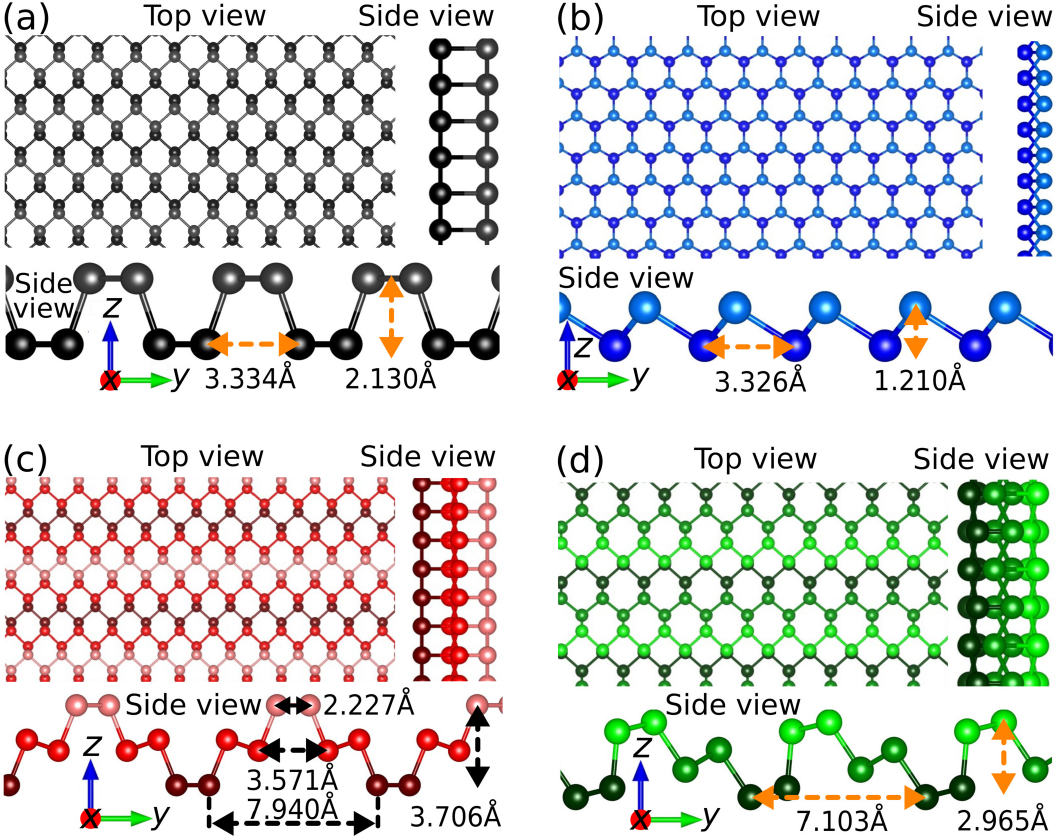}
    \end{center}
    \end{minipage} 
\caption{
Schematic of the structures of (a) black, (b) blue, (c) red and (d) green phosphorenes from different directions. These structures have been illustrated by using the VESTA program~\!\cite{momma2011vesta}.
}
\label{fig:geometry}
\end{figure*}

Black phosphorus, which is a layered semiconducting allotrope of phosphorus, has been synthesized 
first by Bridgman in 1914~\!\cite{bridgman1914two}. This allotrope exhibits high carrier mobility~\!
\cite{warschauer1963electrical}. Using mechanical exfoliation, the black phosphorene was isolated 
by Liu {\it et al.} in 2014~\!\cite{liu2014phosphorene}. Phosphorene, a monolayer of black phosphorus, 
unlike graphene, has a nonzero fundamental band gap that can be engineered by tuning the strain and 
also by the number of layers in a stack~\!\cite{carvalho2016phosphorene,liu2014phosphorene}. The schematic 
of black phosphorene is presented in Fig.~\!\ref{fig:geometry}(a). The crystal structure of black phosphorene is orthorhombic with space group Cmca (No.~\!64). Its lattice constants are $a = 3.3136 \;\text{{\normalfont\AA}}$, $b = 4.3763 \;\text{{\normalfont\AA}}$ and $c = 10.478\; \text{{\normalfont\AA}}$. Four phosphorus atoms are in its crystal cell, but only one of them is inequivalent and located at the position of 8f $(0, 0.08056, 0.10168)$. In our simulations, the black phosphorene sheet sizes are considered as $L_{x}=43.076 \; \text{{\normalfont\AA}}$ and $L_y = 43.763 \;\text{{\normalfont\AA}}$.

Then, the next phosphorus allotrope, which is the blue one, was proposed by Zhu and Tom\'{a}nek~\!\cite{zhu2014semiconducting} by changing the structure of the black phosphorene to a hexagonal structure (graphene-like). Blue phosphorene, unlike the black one, has indirect and wide fundamental band gap. Its crystal structure is hexagonal lattice with P-3m1 (No.~\!164) space-group symmetry and its lattice constants are $a = b = 3.326 \; \text{{\normalfont\AA}}$ and $ c = 17.0 \; \text{{\normalfont\AA}}$. Only one inequivalent phosphorus atom exists in the unit cell, occupying the atomic position of $(0.667, 0.333, 0.536)$. The schematic of blue phosphorene is illustrated in Fig.~\!\ref{fig:geometry}(b). The values of $L_{x}$ and $L_{y}$ of the blue phosphorene sheet in our MD simulations are 43.238 $\text{{\normalfont\AA}}$ and 46.086 $\text{{\normalfont\AA}}$, respectively.

In 2015, Zhao {\it et al.}~\!\cite{zhao2015new} have predicted red phosphorene structure by restructuring 
the segments of black and blue phosphorenes. The red phosphorene has a rectangular lattice with Pbcm (No.~\!57) space group symmetry, and its lattice constants are $a = 3.29\; \text{{\normalfont\AA}} $, 
$b = 9.27\; \text{{\normalfont\AA}}$ and $c = 17.00\; \text{{\normalfont\AA}}$. Eight phosphorus atoms
are in its crystal cell, but only two of them are inequivalent, located at positions of $(0.750, 0.831, 0.391)$ and $(0.750, 0.921, 0.515)$. Fig.~\!\ref{fig:geometry}(c) shows the schematic of the red phosphorene with sheet sizes $L_{x}=42.77 \; \text{{\normalfont\AA}}$ and $L_{y}=46.350 \; \text{{\normalfont\AA}}$ in our MD simulations.

Later, in 2017 the green phosphorene was predicted by Hang {\it et al.}~\!\cite{han2017prediction}. It is constructed from the black phosphorene by flipping every twelfth row of bonds upside down, followed by the dislocation of armchair ridges after every fourth row. The green phosphorene has monoclinic space group with C2/m  symmetry (No.~\!12) and has the equilibrium lattice parameters of $a =10.492\; \text{{\normalfont\AA}} $, $b = 3.312\; \text{{\normalfont\AA}} $, $c = 7.894\; \text{{\normalfont\AA}} $, $\alpha = 90°$, $\beta = 59.44°$ and $\gamma =90°$ as well as three inequivalent Wyckoff positions of 4i $(0.768, 0, 0.893)$, 4i $(0.588, 0, 0.541)$ and 4i $(0.017, 0, 0.760)$~\!\cite{kaewmaraya2021novel}. Panel (d) in Fig.~\!\ref{fig:geometry} presents the schematic of the green phosphorene. The size of the green phosphorene sheet in our MD simulations are $L_x=39.7401 \; \text{{\normalfont\AA}}$ and $L_y=41.451 \;\text{{\normalfont\AA}}$.

To create a nanochannel, two parallel phosphorene sheets are located at a distance $H$ from each other. Since the surfaces of sheets are rough, the corresponding mean height, $\bar{h}^{j}$, of the top and bottom sheets are calculated as
\begin{equation}
  \bar{h}^{j}=\dfrac{1}{N}{\displaystyle\sum_{i=1} ^{N} {z^{j}_i}}\;,
\end{equation}
where $j=$ top and bottom denote to the top and bottom sheets, respectively, $N$ is the number of atoms in the corresponding sheet and $z^{j}_i$ is the $z-$coordinate of the $i$th atom on the sheet $j$. Then, the height of channel, which is the mean distance between two parallel sheets, is obtained by $H=\bar{h}^{\textrm{top}}-\bar{h}^{\textrm{bottom}}$. In all of our MD simulations the value of the channel height is fixed at $H= ~\! 50 ~\!\text{{\normalfont\AA}}$. The surface roughness of phosphorene sheets is calculated by
\begin{equation}
  R_{\textrm{s}}^j = \sqrt{\dfrac{1}{N}{\displaystyle\sum_{i=1} ^{N} {(z^{j}_{i}-\bar{h}^{j})^2}}}\;.
\end{equation}
The values of the surface roughness for black, blue, red and green phoshphorene sheets are $R_{\textrm{s}}= $ 1.0654, 0.6060, 1.3238 and 1.0939, respectively.
The roughness ratio $R_\textrm{r}$ is another important parameter of a rough surface, that is defined as 
\begin{equation}
R_{\textrm{r}} = A_{\textrm{act}}/(L_x \times L_y) \;,
\end{equation}
where $A_{\textrm{act}}$ is the actual surface area. The value of the roughness ratio for black, blue, red and green phosphorene sheets are $R_\textrm{r} = $ 2.0419, 1.5762, 1.9465 and 1.9436, respectively. Moreover, the electrical surface charge density $\sigma$ is obtained by
\begin{equation}
\sigma=\frac{1}{2A_{\textrm{act}}} (N_{\textrm{counterion}}-N_{\textrm{co-ion}}) Q_{\textrm{e}} \;,
\label{sigma_eq}
\end{equation}
where $Q_{\textrm{e}}$ is the electron charge, $N_{\textrm{counterion}}$ and $N_{\textrm{co-ion}}$ are the number of counterions and co--ions in the solution, respectively.

%%%%%%%%%%%%%%%%%%%%%%%%%%%%%%%%%%%%%%%%%%%%%%%%%%%%%%%%%%%%%%%%%%%%%%%%%%%%%%%%%%%%%%%%%%%%%%%%%%%%%%
%%%%%%%%%%%%%%%%%%%%%%%%%%%%%%%%%%%%%%%%%%%%%%%%%%%%%%%%%%%%%%%%%%%%%%%%%%%%%%%%%%%%%%%%%%%%%%%%%%%%%%

\section{MOLECULAR DYNAMIC SIMULATIONS} \label{MD}

\begin{figure*}[t]
	\begin{minipage}{1.0\textwidth}
    \begin{center}
        \includegraphics[width=0.8\textwidth]{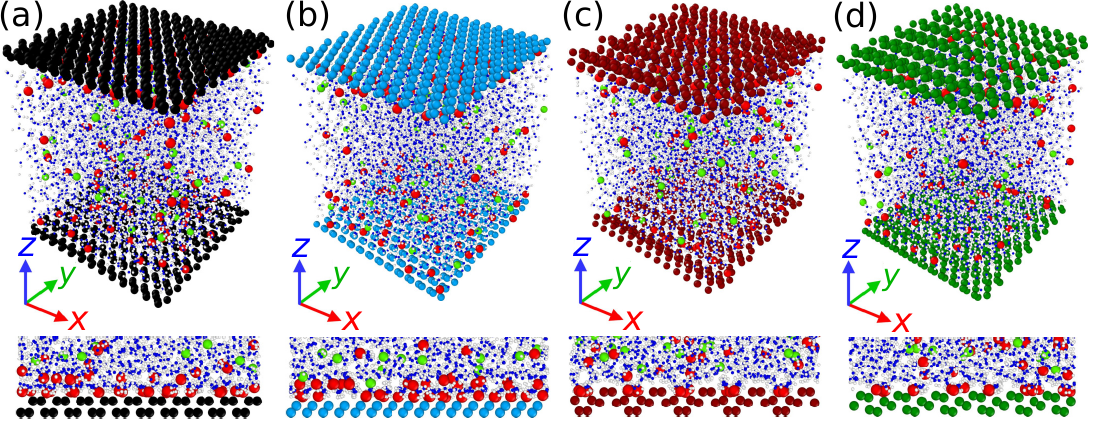}
    \end{center}
    \end{minipage} 
\caption{
(a) Schematic of a neutral system consist of aqueous NaCl solution confined in a nanochannel with two similar parallel black phosphorene walls. Panels (b), (c), and (d) are the same as panel (a) but for different blue, red and green phosphorenes, respectively, as the nanochannel walls.
}
\label{fig:schematic}
\end{figure*}

We consider a three-dimensional model of a system consisting of an aqueous NaCl solution confined between two similar phosphorene walls for four different allotropes. The schematic of the simulation domains, including black, blue, red, and green phosphorenes, are illustrated in panels (a), (b), (c), and (d) in Fig.~\!\ref{fig:schematic}, respectively. To clarify the roughness on the walls, the cross-section of the system has been magnified in the $x$--direction and depicted in the bottom picture in each panel in Fig.~\!\ref{fig:schematic}. The channel average height $H  = 5$~\!nm was carefully chosen to avoid any electric double layer (EDL) overlap at specified ionic concentration and surface charge density. In our simulations, the phosphorus atoms are constrained at their initial positions to maintain a cold wall behavior, while the remaining particles inside the simulation box are free to move. The surface charges are induced by uniformly distributing point charges on phosphorus atoms. The background salt concentration is set to $C_0=1.0$~\!M equally for both Na$^+$ and Cl$^-$ ions that are distributed randomly in the solvent. Total negative charges on the walls are balanced with an excess of Na$^+$ ions adjusted to ensure overall charge neutrality. It should be mentioned that the ion concentration inside the channel is calculated similarly to that of a bulk system, e.g., for a one molar solvent, the number of water is 56 times more than the number of NaCl molecules.

The water is modeled by the extended simple point charge (SPC/E)~\!\cite{SPCE} to describe the water--water interaction. This model of water is commonly used for aqueous NaCl solution and exhibits good performance in predicting hydrodynamic properties of bulk water at various thermodynamic conditions. Using SHAKE algorithm, the water molecules are kept rigid~\!\cite{SHAKE-ryckaert1977numerical}. As the results of other studies~\!\cite{dang1991ion,brodholt1998molecular}, in which the Lennard--Jones (LJ) potential has been used to model systems composed of aqueous sodium chloride solutions and water modeled by the SPC/E have successfully explained the experimental data, the other particles of the system in the present study, i.e. ions in the solution and atoms on the walls are modeled as charged LJ atoms.
The intermolecular interactions between all pairs of particles are modeled by the classical LJ potential as
\begin{equation}
  \label{eq:1}
 V_{\textrm{LJ}}(r_{ij})=4\varepsilon_{ij} \bigg[ \bigg( \frac{\sigma_{ij}}{r_{ij}} \bigg)^{12} - \bigg( \frac{\sigma_{ij}}{r_{ij}} \bigg)^6 \bigg],  \qquad     r \leq r_{\textrm{c}}
\end{equation}
where $\sigma_{ij}$ and $\varepsilon_{ij}$ are the effective LJ size of particles and the depth of the potential well, respectively, and $r_{ij}$ is the separation distance between two atoms. The LJ potential is truncated and shifted at a cutoff length equal to $r_{\textrm{c}} = 1.1$~\!nm. The interaction parameters of the ions and phosphorus atoms are chosen according to the Refs.~\!\citenum{smith1994computer} and \citenum{zhang2016molecular}, respectively. For the LJ parameters between dissimilar atoms the Lorentz--Berthelot combination has been used as 
$ \varepsilon_{ij} = \sqrt{\varepsilon_{i}\varepsilon_{j}}$ and $\sigma_{ij} = (\sigma_i+\sigma_j)/2$. 
As the mass and size of the hydrogen atoms are much smaller than those of the oxygen atom, the hydrogen atoms contributions to the LJ interactions are negligible~\!\cite{Wang2011non}. Consequently, only the oxygen atoms contributions have been taken into account to obtain the LJ interactions between water molecules. The interaction parameters of atomic species are presented in Table~\!\ref{tbl:parameters}.

\begin{table}[]
  \centering
  \caption{Interaction parameters of atomic species}
  \label{tbl:parameters}
\begin{tabular}{|c|c|c|c|}
	\hline
\text{Atoms}    & $\sigma(\textrm{\AA})$ & $\varepsilon(\textrm{Kcal/mol})$ & q($\bar{\textrm{e}}$)   \\\hline
$\textrm{H}$    & 0        & 0       & 0.424  \\\hline
$\textrm{O}$    & 0.1533   & 3.166   & -0.848 \\\hline
$\textrm{Na}^+$ & 0.0148   & 2.575   & 1      \\\hline
$\textrm{Cl}^-$ & 0.106    & 4.448   & -1    \\\hline
$\textrm{P}$    & 0.400    & 3.330   & \text{varying}       \\\hline
\end{tabular}
\end{table}

The electrostatic interaction between atoms $i$ and $j$ is given by the Coulomb potential as
\begin{equation}
  \label{eq:4}
  V_{\textrm{Coulomb}} (r_{ij})=\frac{1}{4\pi\varepsilon_0}  \frac{q_i q_j}{r_{ij}} ,
\end{equation}
where $q_i$ and $q_i$ are the charges on atoms $i$ and $j$, respectively, and $\varepsilon_0$ is vacuum permittivity. To speed up the computation of the long-range electrostatic interactions, the particle--particle--particle--mesh (PPPM) algorithm with a root mean accuracy of $10^{-4}$, is employed~\!\cite{hockney1989particle}. All simulations are performed using Large--scale Atomic/Molecular Massively Parallel Simulator (LAMMPS)~\!\cite{LAMMPS, Brown11, Brown12} package. Periodic boundary conditions are applied in the $x$ and $y$ directions, while a slab modification is applied to calculate the electrostatic interaction for the reduced periodicity due to confinement in the $z$--direction~\!\cite{yeh1999ewald}.
To integrate the equations of motion, the Verlet algorithm with the time step of 2~\!fs is used. Before applying an electric field, the system is thermally equilibrated using a NVT ensemble for 2~\!ns. To this end, initial velocities of water molecules and ions are randomly assigned using uniform distribution function at temperature 300~\!K. We use Nose--Hoover thermostat to maintain a constant temperature. 
Electroosmotic flow is induced by an external electric field applied to the fluid in the $x$ or $y$ directions. 
Temperature is maintained constant using a velocity rescale algorithm, where only the velocity components in the directions orthogonal to the flow, i.e., $y$ and $z$ components, are adjusted. The applied electric field is constant and equal to $0.25$~\!V/nm. Due to the dielectric breakdown of water, using high electric fields in experimental conditions is challenging, but it is required in MD simulation to minimize thermal noise. We run each flow simulation for 10~\!ns to ensure that the flow reaches its steady state. For statistical averaging, the simulations have been performed for an additional 50~\!ns. The results are collected using about 100 equally spaced slab-bins (parallel to the $x$--$y$ plane) in the $z$ direction, which provides a proper resolution for density and velocity profiles.

%%%%%%%%%%%%%%%%%%%%%%%%%%%%%%%%%%%%%%%%%%%%%%%%%%%%%%%%%%%%%%%%%%%%%%%%%%%%%%%%%%%%%%%%%%%%%%%%%%%%%%
%%%%%%%%%%%%%%%%%%%%%%%%%%%%%%%%%%%%%%%%%%%%%%%%%%%%%%%%%%%%%%%%%%%%%%%%%%%%%%%%%%%%%%%%%%%%%%%%%%%%%%

\section{RESULTS} \label{results}

%%%%%%%%%%%%%%%%%%%%%%%%%%%%%%%%%%%%%%%%%%%%%%%%%%%%%%%%%%%%%%%%%%%%%%%%%%%%%%%%%%%%%%%%%%%%%%%%%%%%%%
%%%%%%%%%%%%%%%%%%%%%%%%%%%%%%%%%%%%%%%%%%%%%%%%%%%%%%%%%%%%%%%%%%%%%%%%%%%%%%%%%%%%%%%%%%%%%%%%%%%%%%

\subsection{Electric double layer} \label{EDL}
Inside the nanochannel, a charged region is formed near the channel wall, called the electric double layer (EDL). The EDL consists of two parallel layers: (a) a Stern layer with a typical thickness of one ion diameter that forms next to the wall, and (b) a diffuse layer formed between the Stern layer and the ionic bulk of the system. The ionic bulk of the system is the region wherein the concentrations of co-ions and counterions are almost the same. Due to the very strong electrostatic forces on the Stern layer, the concentration of co--ions ($\textrm{Cl}^-$) in this layer is negligible, and the counterions ($\textrm{Na}^{+}$) in this layer are almost immobile. By attracting a definite number of counterions, the Stern layer is saturated due to the combination of the lack of enough space in the vicinity of the nanochannel wall and the repulsive Coulombic interactions between counterions. The rest of the attracted counterions accumulate in the diffuse layer. Those counterions attracted in the Stern layer shield a part of the surface electric charges. Therefore, the net effective electrostatic force acting on each particle within the diffuse layer is weaker than that in the Stern layer. By applying an external electric field, the ions within the diffuse layer can be able to move, that in turn affects the electroosmotic flow. In fact, the Stern layer, in turn affects the characteristics of the diffuse layer, and the diffuse layer can control the electroosmotic flow. Therefore, the role of the Stern layer as an effective factor in electroosmotic flow is very important. It should be mentioned that the Stern layer is not well defined in theory, and the Poisson--Boltzmann equation describes ion distribution only in the diffuse layer. One of the important advantages of the MD simulation is that it can separately take into account the contributions of both the Stern layer as well as the diffuse layer on the electroosmotic flow, which are distinguished by comparing the co-ions and counterions distribution profiles in the Stern as well as in the diffuse layers. Therefore, using the MD simulation, one can make a detailed analysis of different contributions to the EDL.

\begin{figure*}[t]
	\begin{minipage}{1.0\textwidth}
    \begin{center}
        \includegraphics[width=1.0\textwidth]{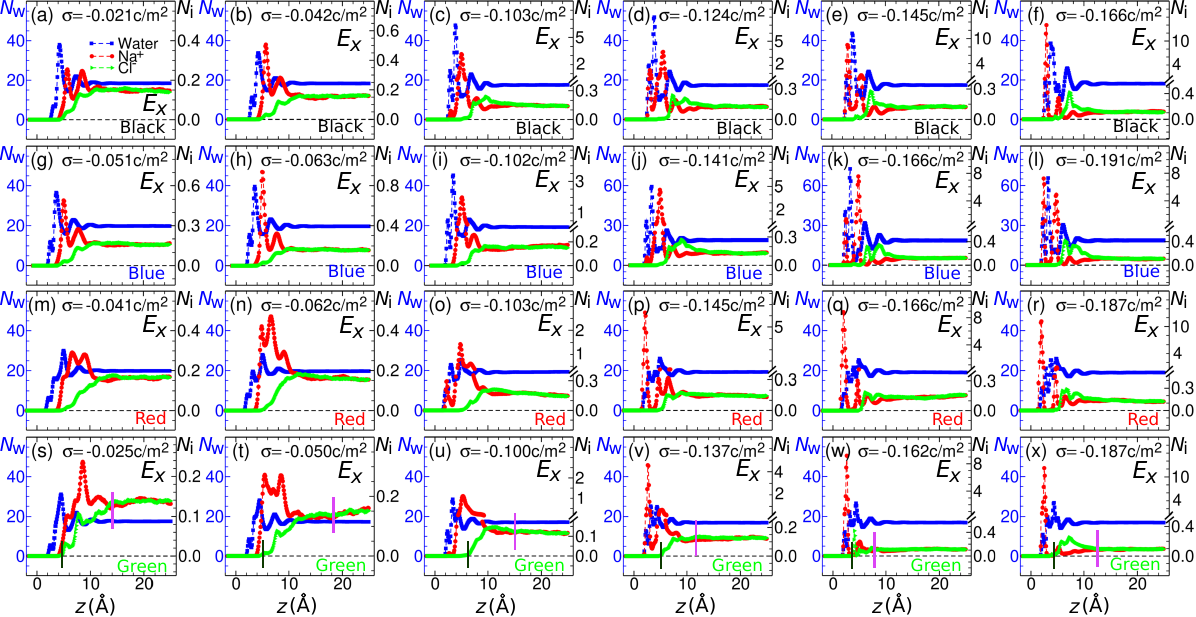}
    \end{center}
    \end{minipage} 
\caption{
(a) The number of water molecules (in blue color), and the number ions $\textrm{Na}^{+}$ (in red color) and $\textrm{Cl}^{-}$ (in green color) as a function of $z$ for a fixed value of the surface charge density $\sigma = -0.021$~\!C/m$^2$ for black phosphorene as the nanochannel walls. The external electric field is applied in the $x$ direction. Panels (b)--(f) are the same as panel (a) but for different values of the surface charge density $\sigma = -0.042$~\!C/m$^2$ to $-0.166$~\!C/m$^2$, respectively.
Panels (g)--(l), (m)--(r), and (s)--(x) are similar to panels (a)--(f) but for different phosphorene allotropes of blue, red, and green, respectively. The locations of Stern layer and diffusive layer have been denoted by short dark green and pink lines in (s)--(w) panels.
}
\label{fig:EDL_Ex}
\end{figure*}

\begin{figure*}[t]
	\begin{minipage}{1.0\textwidth}
    \begin{center}
        \includegraphics[width=1.0\textwidth]{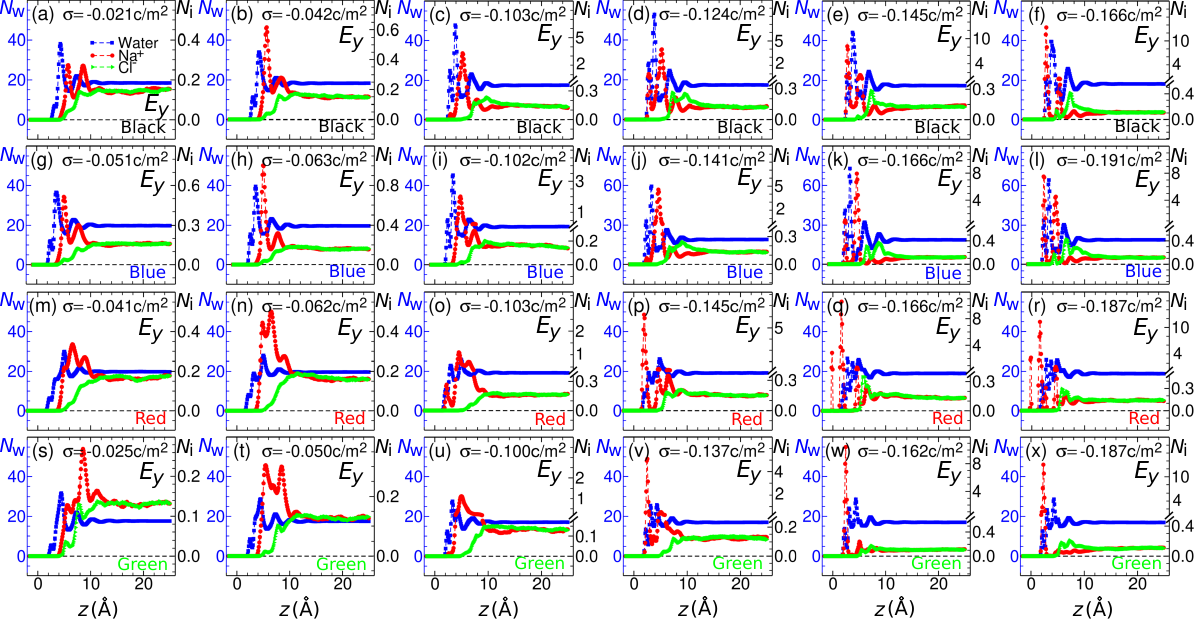}
    \end{center}
    \end{minipage} 
\caption{
(a) The number of water molecules (in blue color), and the number ions $\textrm{Na}^{+}$ (in red color) and $\textrm{Cl}^{-}$ (in green color) as a function of $z$ for a fixed value of the surface charge density $\sigma = -0.021$~\!C/m$^2$ for black phosphorene as the nanochannel walls. The external electric field is applied in the $y$ direction. Panels (b)--(f) are the same as panel (a) but for different values of the surface charge density $\sigma = -0.042$~\!C/m$^2$ to $-0.166$~\!C/m$^2$, respectively. 
Panels (g)--(l), (m)--(r), and (s)--(x) are similar to panels (a)--(f) but for different phosphorene allotropes of blue, red, and green, respectively.
}
\label{fig:EDL_Ey}
\end{figure*}

In this subsection, the Stern layer, which is formed in the vicinity of the nanochannel walls, is studied. To this end, the effect of different phosphorene allotropes with various electric surface charge densities on the characteristics of the Stern layer is discussed. To quantitatively introduce the Stern layer, the system has been divided into very thin rectangular cube slabs, each with sizes of $L_x$, $L_y$ and $\delta z = 0.1$~\!$\text{{\normalfont\AA}}$ and to have reasonable accuracy for the desired quantities it has been averaged over 100 different uncorrelated snapshots of the system. Then, in Fig.~\!\ref{fig:EDL_Ex} the number of water molecules (in blue color), the number of ions $\textrm{Na}^{+}$ (in red color) and $\textrm{Cl}^{-}$ (in green color) in each thin slab have been plotted as a function of $z$ (the distance between the corresponding slab and the bottom nanochannel wall) for various surface charge densities $\sigma = -0.021$~\!C/m$^2$ [depicted in panel (a)] to $\sigma = -0.166$~\!C/m$^2$ [shown in panel (f)] for black phosphorene as the channel walls. The external electric field is applied to the fluid in the $x$ direction. As seen from panel (a) to (f) with the increase in the surface charge density (absolute value), the height of the first peak in the number profile of the counterions $\textrm{Na}^{+}$ (in red color) increases too, and it occurs in the smaller values of $z$.The reason is that with the increase in the negative electrical charges on the phosphorene wall, more $\textrm{Na}^+$ ions are attracted to the wall, and due to the volume excluded repulsion fewer water molecules stay in the vicinity of the wall. To present this in more detail, the number of water molecules profile (in blue color) is plotted in panels (a)--(f) in Fig.~\!\ref{fig:EDL_Ex}. Due to the hydration of the ions, for small values of $|\sigma|$ [panels (a) and (b)] the location of the first peak in the water molecules number profile (in blue color) is closer to the wall than that of the first peak for the counterions profile ($\textrm{Na}^+$ in red color). However, as mentioned above, with the increase in the absolute value of the surface charge density, the peaks of the number profile for $\textrm{Na}^+$ move closer to the wall. Similar results have been obtained for other allotropes of phosphorene, i.e., blue [panels (g)--(l)], red [panels (m)--(r)], and green [panels (s)--(x)] phosphorenes, when the external electric field is applied in the $x$ direction on the fluid. As seen, the results for other allotropes of phosphorene are similar to those of black phosphorene. Moreover, similar results for the number profiles have been obtained for the case, wherein the external electric field is applied in the $y$ direction as depicted in Fig.~\!\ref{fig:EDL_Ey}.

\begin{figure}[t]
	\begin{minipage}{0.5\textwidth}
    \begin{center}
    		\hspace{-0.5cm}
        \includegraphics[width=0.9\textwidth]{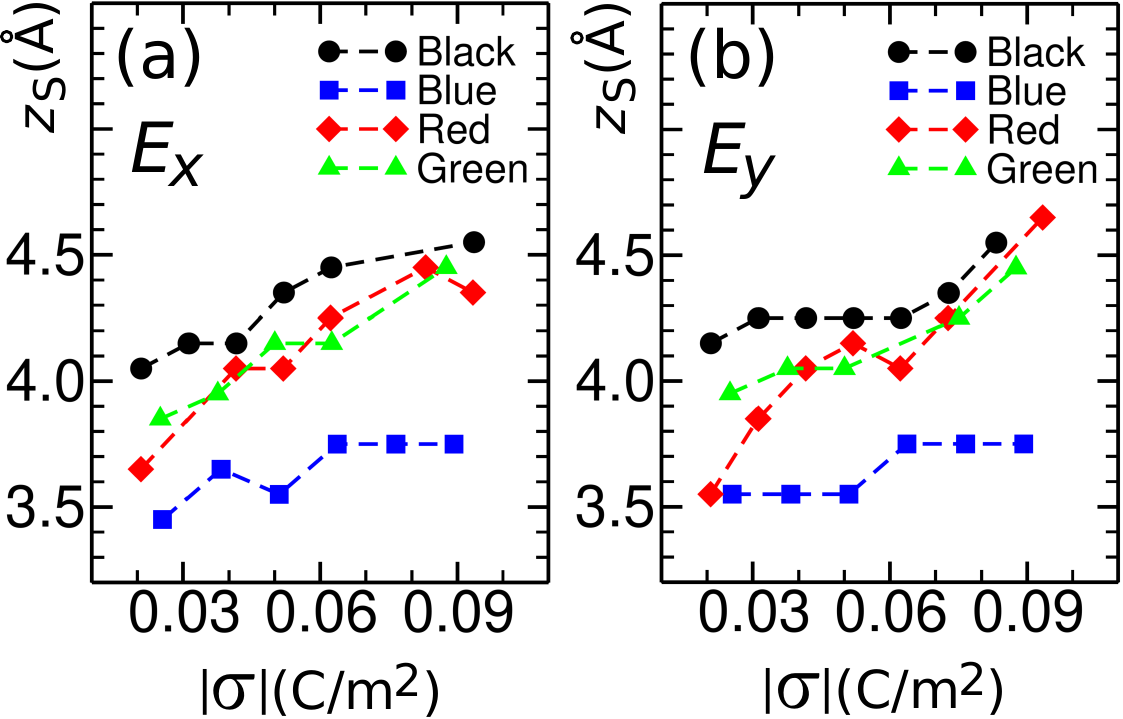}
    \end{center}
    \end{minipage} 
\caption{
(a) The thickness of Stern layer as function of $|\sigma|$ for different black (in black color), blue (in blue color), red (in red color), and green (in green color) phosphorene allotropes, when the external electric field is applied in the $x$ direction. Panel (b) is the same as panel (a) but for the case in which the external electric field is applied in the $y$ direction.
}
\label{fig:SL}
\end{figure}

Next, the thickness of the Stern layer $z_{\textrm{S}}$ is investigated for all four allotropes. The Stern layer is the region in the vicinity of the nanochannel wall, wherein there are not any $\textrm{Cl}^-$ co-ions. Using the number profile of $\textrm{Cl}^-$ co-ions $N_i$ (in green color) in panels (a)--(f) in Fig.~\!\ref{fig:EDL_Ex}, the thickness of the Stern layer for the black phosphorene is obtained as a function of the absolute value of the surface charge density $|\sigma|$ when the external electric field is applied in the $x$ ($E_x$) and $y$ ($E_y$) directions. The same procedure has been repeated for other allotropes of phosphorene, i.e. blue, red and green ones, and the results are presented in panels (a) and (b) in Fig.~\!\ref{fig:SL} for $E_x$ and $E_y$, respectively. The results in both panels show that at constant absolute value of surface charge density $|\sigma|$, the thicknesses of the Stern layer $z_{\textrm{S}}$ for the blue and black phosphorenes have the smallest and largest values, respectively. The reason is that according to Eq.~\!(\ref{sigma_eq}) at constant $|\sigma|$ on the surface with a greater roughness ratio, there exists more electric charges, and due to the neutrality condition, the system contains more counterions. As the number of surface charges is greater, the number of attracted counterions to the surface is larger, too, due to the stronger electrostatic interactions between the surface charges and the counterions. Therefore, the Stern layer thickness is larger for the black phosphorene. For the blue phosphorene, as the roughness ratio is the smallest one, the number of attracted counterions to the surface is the smallest. On the other hand, according to the above scenario for each individual allotrope as the value of $|\sigma|$ increases the number of attracted counterions to the surface increases and consequently, the Stern layer thickness increases, too. 
In addition, the Stern layer thicknesses $z_{\textrm{S}}$ for the red and green phosphorenes are in between those of the black and blue ones in both panels (a) and (b). This can be explained by the fact that the value of the roughness ratio of the red (1.9465) and green (1.9436) phosphorenes are between those of the blue (1.5762) and black (2.0419) ones. 
Moreover, the surface roughness and groove width (see Fig.~\!\ref{fig:geometry}) play additional important roles. As the groove widths of red (7.940$\textrm{\AA}$) and green (7.103$\textrm{\AA}$) phosphorenes are larger than that of black one (3.33$\textrm{\AA}$), more counterions can enter and stay in the space inside the grooves of red and green phosphorenes than the black one. Consequently, the co-ions exist closer to the wall in the channels made by red and green phosphorenes. This to some extent yields a reduction in the thickness of the Stern layer for the red and green phosphorenes compared to the black one.

The results in panels (a) and (b) in Fig.~\!\ref{fig:SL} are summarized in Tables~\!\ref{tab:Zs-Ex} and \!\ref{tab:Zs-Ey}, respectively.

\begin{table*}[]
		\begin{tabular}{|cc|cc|cc|cc|}
			\hline
			\multicolumn{2}{|c|}{BLACK}                                                                                                                        & \multicolumn{2}{c|}{BLUE}                                                                                                                         & \multicolumn{2}{c|}{RED}                                                                                                                          & \multicolumn{2}{c|}{GREEN}                                                                                                                        \\ \hline
			\multicolumn{1}{|c|}{$\sigma$($\textrm{C}/{\textrm{m}^2}$)} & \begin{tabular}[c]{@{}c@{}}$\textrm{z}_{\textrm{S}}$\\ $(\textrm{\AA})$\end{tabular} & \multicolumn{1}{c|}{$\sigma$($\textrm{C}/{\textrm{m}^2}$)} & \begin{tabular}[c]{@{}c@{}}$\textrm{z}_{\textrm{S}}$\\ $(\textrm{\AA})$\end{tabular} & \multicolumn{1}{c|}{$\sigma$($\textrm{C}/{\textrm{m}^2}$)} & \begin{tabular}[c]{@{}c@{}}$\textrm{z}_{\textrm{S}}$\\ $(\textrm{\AA})$\end{tabular} & \multicolumn{1}{c|}{$\sigma$($\textrm{C}/{\textrm{m}^2}$)} & \begin{tabular}[c]{@{}c@{}}$\textrm{z}_{\textrm{S}}$\\ $(\textrm{\AA})$\end{tabular} \\ \hline
			\multicolumn{1}{|c|}{0.021}                                 & 4.05                                                                                 & \multicolumn{1}{c|}{0.026}                                 & 3.45                                                                                 & \multicolumn{1}{c|}{0.021}                                 & 3.65                                                                                 & \multicolumn{1}{c|}{0.025}                                 & 3.85                                                                                 \\ \hline
			\multicolumn{1}{|c|}{0.031}                                 & 4.15                                                                                 & \multicolumn{1}{c|}{0.038}                                 & 3.65                                                                                 & \multicolumn{1}{c|}{0.042}                                 & 4.05                                                                                 & \multicolumn{1}{c|}{0.038}                                 & 3.95                                                                                 \\ \hline
			\multicolumn{1}{|c|}{0.042}                                 & 4.15                                                                                 & \multicolumn{1}{c|}{0.051}                                 & 3.55                                                                                 & \multicolumn{1}{c|}{0.052}                                 & 4.05                                                                                 & \multicolumn{1}{c|}{0.05}                                  & 4.15                                                                                 \\ \hline
			\multicolumn{1}{|c|}{0.052}                                 & 4.35                                                                                 & \multicolumn{1}{c|}{0.064}                                 & 3.75                                                                                 & \multicolumn{1}{c|}{0.062}                                 & 4.25                                                                                 & \multicolumn{1}{c|}{0.063}                                 & 4.15                                                                                 \\ \hline
			\multicolumn{1}{|c|}{0.062}                                 & 4.45                                                                                 & \multicolumn{1}{c|}{0.077}                                 & 3.75                                                                                 & \multicolumn{1}{c|}{0.083}                                 & 4.45                                                                                 & \multicolumn{1}{c|}{0.088}                                 & 4.45                                                                                 \\ \hline
			\multicolumn{1}{|c|}{0.094}                                 & 4.55                                                                                 & \multicolumn{1}{c|}{0.089}                                 & 3.75                                                                                 & \multicolumn{1}{c|}{0.093}                                 & 4.35                                                                                 & \multicolumn{1}{c|}{-}                                     & -                                                                                    \\ \hline
		\end{tabular}
		\caption{The thickness of Stern layer as function of $|\sigma|$ for different black, blue, red, and green phosphorene allotropes, when the external electric field is applied in the $x$ direction.}
		\label{tab:Zs-Ex}	
\end{table*}

\begin{table*}[]
		\begin{tabular}{|cc|cc|cc|cc|}
			\hline
			\multicolumn{2}{|c|}{BLACK}                                                                                                                        & \multicolumn{2}{c|}{BLUE}                                                                                                                         & \multicolumn{2}{c|}{RED}                                                                                                                          & \multicolumn{2}{c|}{GREEN}                                                                                                                        \\ \hline
			\multicolumn{1}{|c|}{$\sigma$($\textrm{C}/{\textrm{m}^2}$)} & \begin{tabular}[c]{@{}c@{}}$\textrm{z}_{\textrm{S}}$\\ $(\textrm{\AA})$\end{tabular} & \multicolumn{1}{c|}{$\sigma$($\textrm{C}/{\textrm{m}^2}$)} & \begin{tabular}[c]{@{}c@{}}$\textrm{z}_{\textrm{S}}$\\ $(\textrm{\AA})$\end{tabular} & \multicolumn{1}{c|}{$\sigma$($\textrm{C}/{\textrm{m}^2}$)} & \begin{tabular}[c]{@{}c@{}}$\textrm{z}_{\textrm{S}}$\\ $(\textrm{\AA})$\end{tabular} & \multicolumn{1}{c|}{$\sigma$($\textrm{C}/{\textrm{m}^2}$)} & \begin{tabular}[c]{@{}c@{}}$\textrm{z}_{\textrm{S}}$\\ $(\textrm{\AA})$\end{tabular} \\ \hline
			\multicolumn{1}{|c|}{0.021}                                 & 4.15                                                                                 & \multicolumn{1}{c|}{0.026}                                 & 3.55                                                                                 & \multicolumn{1}{c|}{0.021}                                 & 3.55                                                                                 & \multicolumn{1}{c|}{0.025}                                 & 3.95                                                                                 \\ \hline
			\multicolumn{1}{|c|}{0.031}                                 & 4.25                                                                                 & \multicolumn{1}{c|}{0.038}                                 & 3.55                                                                                 & \multicolumn{1}{c|}{0.031}                                 & 3.85                                                                                 & \multicolumn{1}{c|}{0.038}                                 & 4.05                                                                                 \\ \hline
			\multicolumn{1}{|c|}{0.042}                                 & 4.25                                                                                 & \multicolumn{1}{c|}{0.051}                                 & 3.55                                                                                 & \multicolumn{1}{c|}{0.042}                                 & 4.05                                                                                 & \multicolumn{1}{c|}{0.05}                                  & 4.05                                                                                 \\ \hline
			\multicolumn{1}{|c|}{0.052}                                 & 4.25                                                                                 & \multicolumn{1}{c|}{0.064}                                 & 3.75                                                                                 & \multicolumn{1}{c|}{0.052}                                 & 4.15                                                                                 & \multicolumn{1}{c|}{0.075}                                 & 4.25                                                                                 \\ \hline
			\multicolumn{1}{|c|}{0.062}                                 & 4.25                                                                                 & \multicolumn{1}{c|}{0.077}                                 & 3.75                                                                                 & \multicolumn{1}{c|}{0.062}                                 & 4.05                                                                                 & \multicolumn{1}{c|}{0.088}                                 & 4.45                                                                                 \\ \hline
			\multicolumn{1}{|c|}{0.073}                                 & 4.35                                                                                 & \multicolumn{1}{c|}{0.089}                                 & 3.75                                                                                 & \multicolumn{1}{c|}{0.073}                                 & 4.25                                                                                 & \multicolumn{1}{c|}{-}                                     & -                                                                                    \\ \hline
			\multicolumn{1}{|c|}{0.083}                                 & 4.55                                                                                 & \multicolumn{1}{c|}{0.108}                                 & 4.55                                                                                 & \multicolumn{1}{c|}{0.093}                                 & 4.65                                                                                 & \multicolumn{1}{c|}{-}                                     & -                                                                                    \\ \hline
		\end{tabular}
		\caption{The thickness of Stern layer as function of $|\sigma|$ for different black, blue, red, and green phosphorene allotropes, when the external electric field is applied in the $y$ direction.}
		\label{tab:Zs-Ey}
\end{table*}

%%%%%%%%%%%%%%%%%%%%%%%%%%%%%%%%%%%%%%%%%%%%%%%%%%%%%%%%%%%%%%%%%%%%%%%%%%%%%%%%%%%%%%%%%%%%%%%%%%%%%%
%%%%%%%%%%%%%%%%%%%%%%%%%%%%%%%%%%%%%%%%%%%%%%%%%%%%%%%%%%%%%%%%%%%%%%%%%%%%%%%%%%%%%%%%%%%%%%%%%%%%%%

\subsection{Electroosmotic flow velocity} \label{EFV}

\begin{figure*}[t]
	\begin{minipage}{1.0\textwidth}
    \begin{center}
        \includegraphics[width=0.95\textwidth]{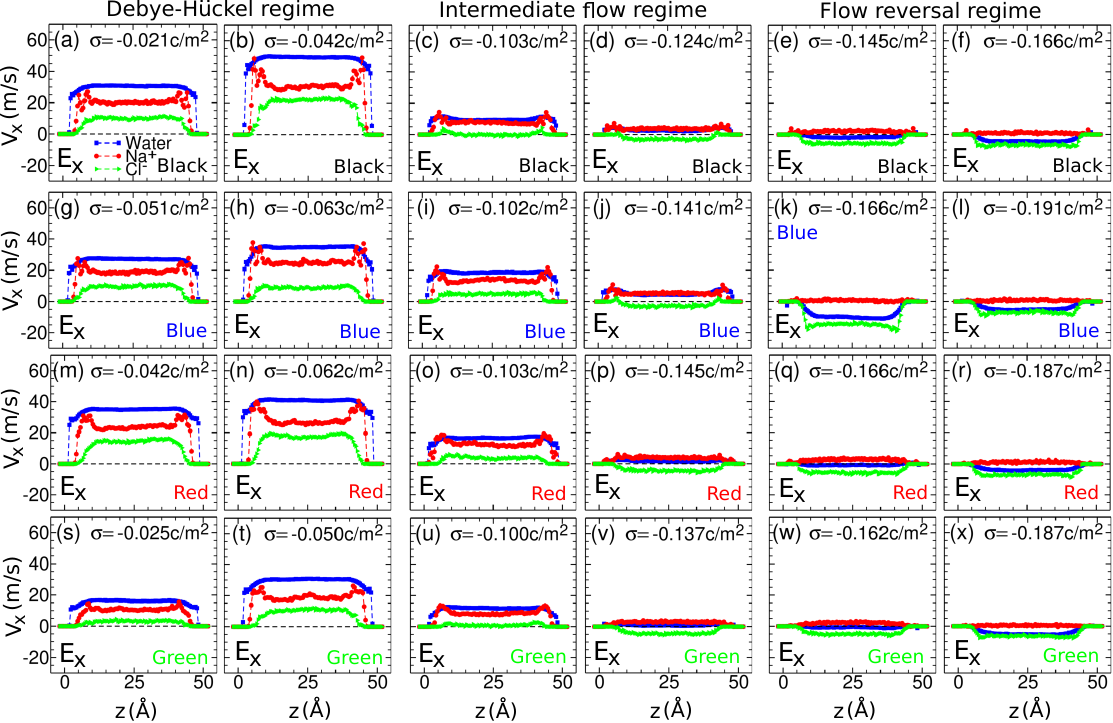}
    \end{center}
    \end{minipage} 
\caption{
(a) The velocity in the $x$ direction $v_x$ in the presence of an external electric field in the $x$ direction as a function of $z$ for a fixed value of surface charge density  $\sigma = -0.021$~\!$\textrm{C}/{\textrm{m}^2}$ for water molecules (in blue color), $\textrm{Na}^+$ counterions (in red color) and $\textrm{Cl}^-$ co-ions (in green color) inside the nanochannel with black phosphorene as its walls. Panels (b)--(f) are the same as (a) but for different values of the surface charge density $\sigma = -0.042$~\!$\textrm{C}/{\textrm{m}^2}$ to -0.166~\!$\textrm{C}/{\textrm{m}^2}$, respectively.
Panels (g)--(l), (m)--(r), and (s)--(x) are similar to panels (a)--(f) but for blue, red, and green phosphorenes as the nanochannel walls.
}
\label{fig:EOF-X-ionic}
\end{figure*}

\begin{figure*}[t]
	\begin{minipage}{1.0\textwidth}
    \begin{center}
        \includegraphics[width=0.95\textwidth]{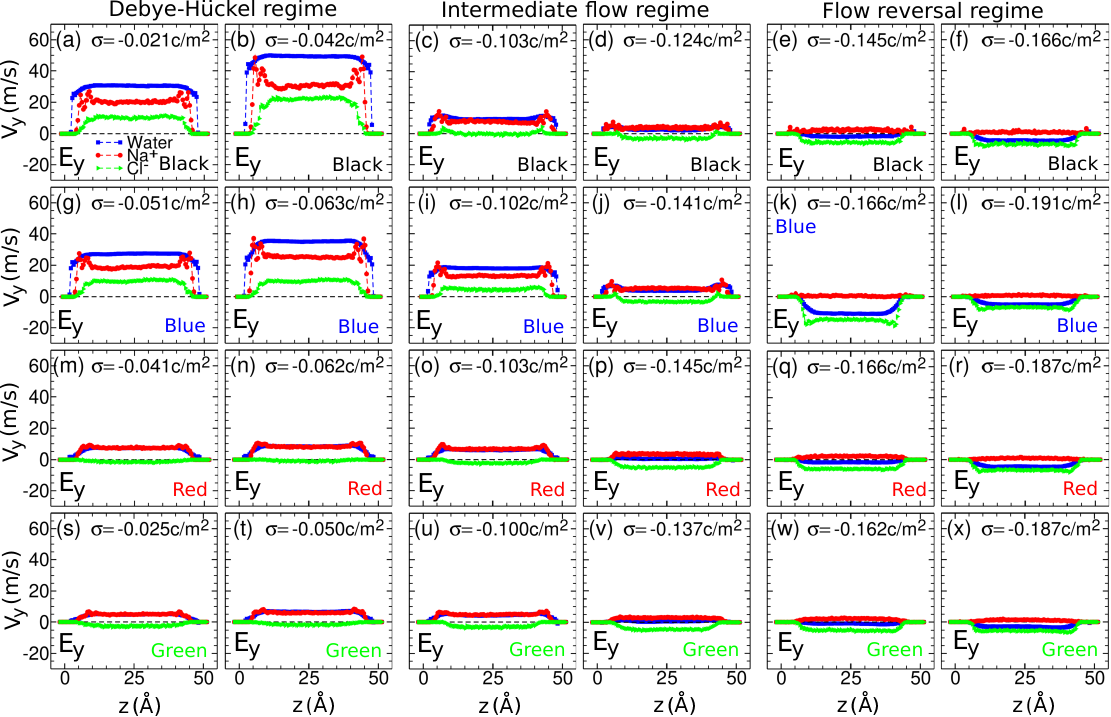}
    \end{center}
    \end{minipage} 
\caption{
(a) The velocity in the $y$ direction $v_y$ in the presence of an external electric field in the $y$ direction as a function of $z$ for a fixed value of surface charge density  $\sigma = -0.021$~\!$\textrm{C}/{\textrm{m}^2}$ for water molecules (in blue color), $\textrm{Na}^+$ counterions (in red color) and $\textrm{Cl}^-$ co-ions (in green color) inside the nanochannel with black phosphorene as its walls. Panels (b)--(f) are the same as (a) but for different values of the surface charge density $\sigma = -0.042$~\!$\textrm{C}/{\textrm{m}^2}$ to -0.166~\!$\textrm{C}/{\textrm{m}^2}$, respectively.
Panels (g)--(l), (m)--(r), and (s)--(x) are similar to panels (a)--(f) but for blue, red, and green phosphorenes as the nanochannel walls.
}
\label{fig:EOF-Y-ionic}
\end{figure*}

In this subsection, the effect of the surface charge density $\sigma$ on the electroosmotic flow velocity is investigated. To this end the value of $\sigma$ is varied in the range $0.01~\!\textrm{C}/{\textrm{m}^2}$ $\leq  \sigma \leq 0.2 $~\!$\textrm{C}/{\textrm{m}^2}$ and the simulations are performed. To find the profile of the electroosmotic velocity in the $x$ direction $v_x$ in the presence of the electric field in the $x$ direction, the system is divided in the $z$ direction to parallel slabs each with a thickness of $0.5~\!\textrm{\AA}$ The mean value of velocity in each slab $v_x$ is obtained by averaging over 15 millions snapshots. 
Different curves of ions velocity profiles in Fig.~\!\ref{fig:EOF-X-ionic}(a)--(f) for black phosphorene as the channel walls can help to understand the physics behind the electroosmotic flow profiles, which is almost the same as the flow for the water molecules (blue curve in each panel).
In Fig.~\!\ref{fig:EOF-X-ionic}(a) for the nanochannel with black phosphorene as its walls, $v_x$ has been plotted as a function of $z$ for fixed value of $\sigma = -0.021$~\!$\textrm{C}/{\textrm{m}^2}$ for water (in blue color), $\textrm{Na}^+$ (counterions, in red color) and $\textrm{Cl}^-$ (co-ions, in green color) molecules. Panels (b)--(f) are the same as (a) but for different values of the surface charge density $\sigma = -0.042$, -0.103, -0.124, -0.145 and -0.166~\!$\textrm{C}/{\textrm{m}^2}$, respectively. 
According to the electroosmotic velocity behavior for various surface charge densities, three different regimes are introduced as Debye--H\"{u}ckel (DH), intermediate and flow reversal. In the DH regime, the electroosmotic flow velocity monotonically increases by increase in the absolute value of the surface charge density within the validity of the DH approximation [see panels (a) and (b) in Fig.~\!\ref{fig:EOF-X-ionic}]. 
Indeed, the Poisson-Boltzmann equation can be linearized for a small zeta potential through a Taylor series expansion, which known as the DH regime~\!\cite{Andelman}. In the present study the DH regime is identified for approximately $\sigma$ $\leq$ 0.05 ${\rm C}/{{\rm m}^2}$, where the PB model is valid.

In panels (a) and (b), which present $v_x$ in the DH regime, the $\textrm{Na}^{+}$ counterions dominate in the diffuse layer and move in the direction of the applying external electric field. Therefore, the counterions drag the electrically neutral water molecules with them. The negatively charged $\textrm{Cl}^{-}$ co-ions tend to move in the opposite direction with respect to the electric field. However, in the DH regime, due to the drag force, they experience a net force in the direction of an electric field, and therefore their motion is in the direction of the electric field with a smaller speed than that of the water flow.
By increasing the absolute value of the surface charge density beyond $\sigma \approx -0.042~\!\textrm{C}/{\textrm{m}^2}$ the overscreening is observed, and the intermediate regime is started, wherein the velocity begins to decrease with increase in the absolute value of the surface charge density, while the flow is still in the direction of the electric field [panels (c) and (d) in Fig.~\!\ref{fig:EOF-X-ionic}]. 
In the intermediate regime, $\textrm{Na}^{+}$ counterions overshield the charged walls. 
%, and the dynamics in the diffuse layer is dominated by the dynamics of the co-ions ($\textrm{Cl}^{-}$). 
Panels (c) and (d) show that the bulk velocity of $\textrm{Cl}^{-}$ co-ions first gradually decreases, and then it begins to move in the opposite direction with respect to the direction of the external electric field. Moreover, by increasing the absolute value of the surface charge density, the velocity of $\textrm{Na}^{+}$ counterions decreases. Indeed, in the intermediate regime, the speed of counterions still overcomes that of the co-ions, which leads to a continuously decrease in the water flow in the direction of the electric field until the flow gets the zero value. 
By increasing the charge density even more, the water velocity gets a negative value, and the flow reversal occurs. The results for the flow reversal regime are presented in panels (e) and (f), in which the flow velocity is in the direction opposite to the external electric field, and the absolute value of flow velocity increases by an increase in the $\sigma$. The reason for the reverse flow is that at very high values of surface charge density due to the very strong interaction between the charged surface and the counterions and in the presence of the external electric field, the effective dynamics in the diffuse layer is dominated by the dynamics of the $\textrm{Cl}^-$ co-ions. Then, the shear stress produced by co-ions in the diffuse layer induces reverse motion to the whole flow.

\begin{figure*}[t]
	\begin{minipage}{1.0\textwidth}
    \begin{center}
    \hspace{-0.5cm}
        \includegraphics[width=0.9\textwidth]{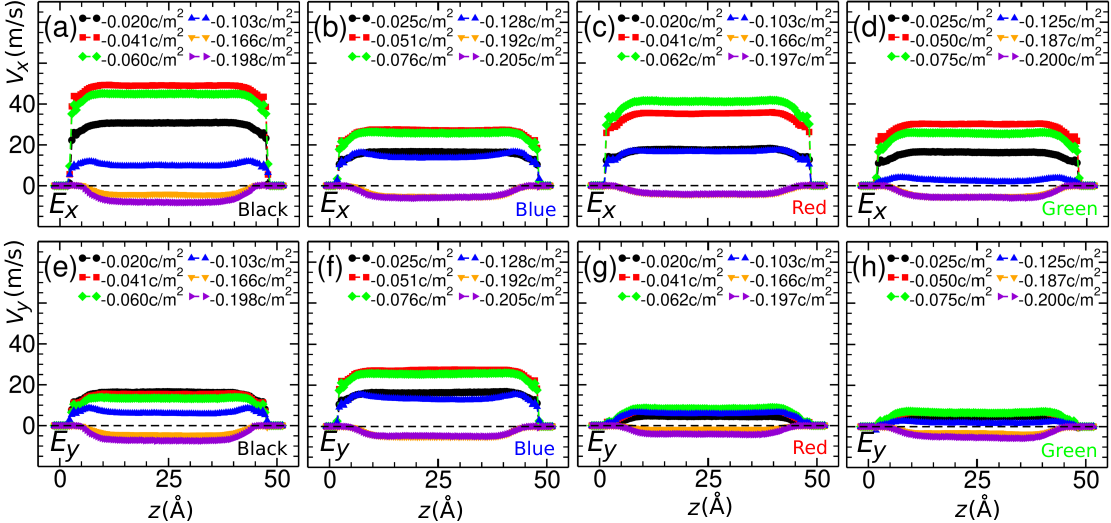}        
    \end{center}
    \end{minipage} 
\caption{
(a) Flow velocity profile $v_x$ as a function of $z$ for black phosphorene for various values of the surface charge density $\sigma = - 0.020~\!\textrm{C}/{\textrm{m}^2}$ (black circles) to $- 0.198~\!\textrm{C}/{\textrm{m}^2}$ (violet triangles left), when the electric field is applied in the $x$ direction. Panels (b)--(d) are similar to panel (a) but for different blue, red and green phosphorenes, respectively.
Panels (e)--(h) are the same as panels (a)--(d) but for showing the flow velocity profiles $v_y$ when the electric field is applied in the $y$ direction.
}
\label{fig:EOF-density}
\end{figure*}

Panels (g)--(l), (m)--(r), and (s)--(x) are similar to panels (a)--(f) but for blue, red, and green phosphorenes as the channel walls, respectively. According to the results in panels (a) to (x) in Fig.~\!\ref{fig:EOF-X-ionic}, similar behavior in the velocity profiles is seen for different phosphorenes. 
Similar data are obtained if the external electric field is applied in the $y$-- direction that can be seen in Fig.~\!\ref{fig:EOF-Y-ionic}.

\begin{figure}[t]
	\begin{minipage}{0.5\textwidth}
    \begin{center}
    \hspace{-0.5cm}
        \includegraphics[width=0.9\textwidth]{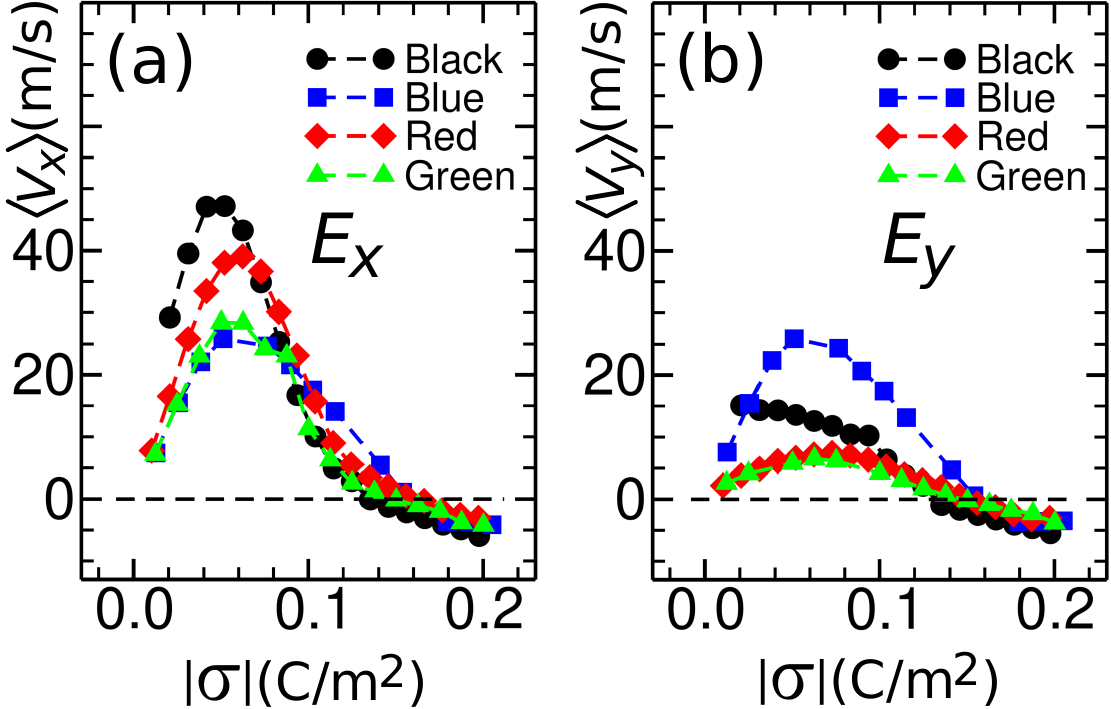}        
    \end{center}
    \end{minipage} 
\caption{
(a) The average flow velocities across the nanochannel $\langle v_x \rangle $ (when the electric field is applied in the $x$ direction) as a function of the absolute value of the surface charge density $|\sigma|$ for black (in black color), blue (in blue color), red (in red color), and green (in green color) phosphorene allotropes. Panel (b) is the same as panel (a) but for $\langle v_y \rangle $ in which the electric field is applied in the $y$ direction.
}
\label{fig:final}
\end{figure}

To compare the flow velocity across the nanochannel for different values of the surface charge density, the flow velocity profiles are plotted as a function of $z$ for black, blue, red, and green phosphorene allotropes in panels (a) to (d) of Fig.~\!\ref{fig:EOF-density}, respectively, when the external applying electric field is in the $x$ direction. Panels (e) to (h) in Fig.~\!\ref{fig:EOF-density} are the same as panels (a) to (d), respectively, but for the case in which the external electric field is in the $y$ direction. 
Finally, the average flow velocities across the nanochannel in the $x$ ($\langle v_x \rangle $) and $y$ ($\langle v_y \rangle $) directions are obtained by averaging over velocities at different slabs depicted in Fig.~\!\ref{fig:EOF-X-ionic} and \ref{fig:EOF-Y-ionic} for the cases in which the electric field is applied in the $x$ and $y$ directions, respectively. In panel (a) of Fig.~\!\ref{fig:final}, the average velocity $\langle v_x \rangle $ has been plotted as a function of the absolute value of the surface charge density $|\sigma|$ for black (in black color), blue (in blue color), red (in red color), and green (in green color) phosphorene allotropes. Panel (b) is the same as panel (a) but for $\langle v_y \rangle $.

The comparison between the results in panel (a) with those of panel (b) of Fig.~\!\ref{fig:final} shows that for the same phosphorene allotrope and a constant value of $|\sigma|$, the flow velocity for the case wherein the external applying electric field is in the $x$--direction [panel (a)] is greater than that of the case wherein the electric field is in the $y$--direction [panel (b)], except for the blue phosphorene case in which the direction of the external electric field does not affect the flow velocity. As the phosphorenes are uniaxial and their roughness are invariant in the $x$ direction, the roughness has a greater effect on the dynamics of the system when the electric field is in the $y$ direction compared to the other case, in which the electric field is in the $x$ direction. Therefore, the effective flow velocity decreases when the electric field is in the direction of $y$. Moreover, as the roughness of blue phosphorene is smaller than the other allotropes and according to the Fig. \ref{fig:schematic}(b), the water molecules are not able to enter into the space inside the grooves of the blue phosphorene, consequently as mentioned above the flow velocities for the blue phosphorene in panels (a) and (b) of Fig.~\!\ref{fig:final} are the same. Finally, three different electroosmotic regimes of DH, intermediate, and reverse can be clearly seen in Fig.~\!\ref{fig:final}. Next, the reverse flow regime is considered in more detail.

%%%%%%%%%%%%%%%%%%%%%%%%%%%%%%%%%%%%%%%%%%%%%%%%%%%%%%%%%%%%%%%%%%%%%%%%%%%%%%%%%%%%%%%%%%%%%%%%%%%%%%
%%%%%%%%%%%%%%%%%%%%%%%%%%%%%%%%%%%%%%%%%%%%%%%%%%%%%%%%%%%%%%%%%%%%%%%%%%%%%%%%%%%%%%%%%%%%%%%%%%%%%%

\subsection{Flow reversal} \label{FR}

\begin{figure}[t]
	\begin{minipage}{0.5\textwidth}
    \begin{center}
    \hspace{-0.5cm}
        \includegraphics[width=0.75\textwidth]{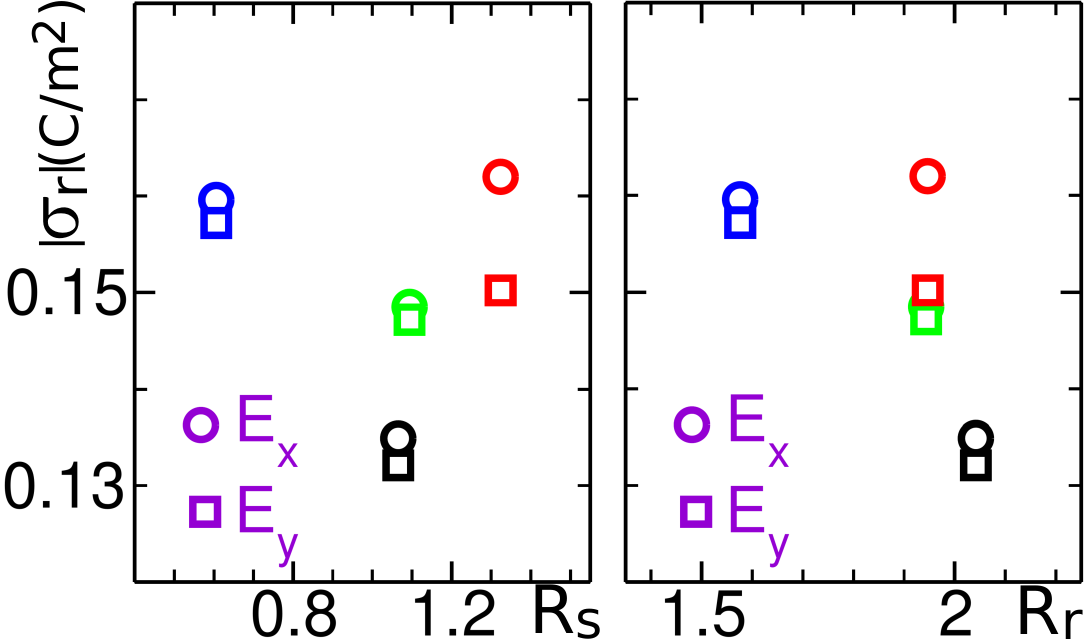}        
    \end{center}
    \end{minipage} 
\caption{(a) The absolute value of the reverse surface charge density $|\sigma_r|$ versus the surface roughness $R_{\textrm{s}}$ for the external electric fields in the $x$ direction ($E_x$, circles) and in the $y$ direction ($E_y$, squares) for different phosphorene allotropes of black (in black color), blue  (in blue color), red (in red color) and green (in green color). Panel (b) is the same as panel (a) but for $|\sigma_r|$ versus the roughness ratio $R_{\textrm{r}}$.
}
\label{fig:roughness}
\end{figure}

In a system composed of highly charged surfaces the excessive counter-ions adsorption in the Stern layer induces the overscreening (charge inversion phenomenon). Indeed, the total adsorbed counter-ions charge on the surface is more than that of the surface. As whole system and its bulk are neutral, and due to the presence of excessive adsorbed counter-ions next to the surface, the co-ions assemble more than counter-ions in the diffusive layer which is located between the Stern layer and the bulk of system. This effectively changes the polarity of EDL.
This is similar to the case where in the system includes positively charged wall (composed of negatively charged surface and positively charged Stern layer) and negatively charged diffusive layer, which leads to form a newly secondary EDL in the vicinity of positively charged wall.
Applying the electric field, the co-ions belonging to in the newly formed diffuse layer start moving in the opposite direction to the electric field. The movement of co-ions in the diffusive layer in opposite direction causes to move the water molecules in same direction that is called ”reversed electroosmotic flow”.\cite{celebi2019molecular,qiao2004charge,rezaei2015surface}
As the absolute value of the surface charge density, $|\sigma|$, increases, at a certain density, $\sigma_r$, the direction of the EOF reverses to the opposite direction with respect to the direction of the external electric field. The values of $\sigma_r$ for all allotropes and for both $x$ and $y$ directions of the external electric field are obtained using the results in Fig.~\!\ref{fig:final}. The surface roughness $R_{\textrm{s}}$, roughness ratio $R_{\textrm{r}}$ and the reverse surface charge density $\sigma_{\textrm{r}}$ of all allotropes are presented in Table~\!\ref{tab:roughness}. The data in Table~\!\ref{tab:roughness} has been depicted in Fig.~\!\ref{fig:roughness}. The results in Fig.~\!\ref{fig:roughness} reveal the fact that the value of $|\sigma_{\textrm{r}}|$ increases with the increase in the surface roughness for both $x$ and $y$ directions, except for the case of the blue phosphorene in which due to the small value of the roughness the dynamics of the system is not sensitive to the direction of the external electric field. 
The increase in $|\sigma_{\textrm{r}}|$ with respect to $R_{\textrm{s}}$ is because as the surface roughness increases, more electric charge on the wall is needed in order to the reverse EOF flow occur. 
According to the results in Fig.~\!\ref{fig:roughness} for all allotropes, the values of the $|\sigma_r|$ for the case in which the electric field in the $x$ direction are larger than those of the cases in which the electric field is in the $y$ direction. 
The reason is that the EOF velocity for the case of an electric field in the $x$ direction ($E_x$) is larger than that of the case in which the electric field is in the $y$ direction ($E_y$). Therefore, to make the flow in the reverse direction, more electric charge density is required for the case of $E_x$ compared to the case of $E_y$. %According to the results in Table~\!\ref{tab:roughness} the reverse flow is more controlled by the surface roughness than the roughness ratio, which is in agreement with the results of Ref.[\!\!\citenum{lu2017electroosmotic}] by Lu {\it et al.} for the EOF inside a nanochannel with rough fractal surfaces.

\begin{table}
	\begin{tabular}{|c|c|c|c|c|}
		\hline
		& Blue & Black & Green & Red \\
		\hline
		$R_{\textrm{s}}$~\!$(\textrm{\AA})$ & 0.606  & 1.065  & 1.094  & 1.324 \\
		\hline
		$R_{\textrm{r}}$  &1.576 & 2.042 & 1.944 & 1.946\\
		\hline
		$\sigma_{\textrm{r}}$~\!\!($\textrm{C}$/$\textrm{m}^2$):$E_x$ & 0.160 & 0.135 & 0.148 & 0.162 \\
		\hline
		$\sigma_{\textrm{r}}$~\!\!($\textrm{C}$/$\textrm{m}^2$):$E_y$ & 0.157 & 0.132 & 0.147 & 0.150 \\
		\hline
		
	\end{tabular}
	\caption{The values of roughness and $\sigma_r$ with electric fields $E_x$ and $E_y$ of four phosphorene allotropes.}
	\label{tab:roughness}	
\end{table}

%%%%%%%%%%%%%%%%%%%%%%%%%%%%%%%%%%%%%%%%%%%%%%%%%%%%%%%%%%%%%%%%%%%%%%%%%%%%%%%%%%%%%%%%%%%%%%%%%%%%%%
%%%%%%%%%%%%%%%%%%%%%%%%%%%%%%%%%%%%%%%%%%%%%%%%%%%%%%%%%%%%%%%%%%%%%%%%%%%%%%%%%%%%%%%%%%%%%%%%%%%%%%

\section{SUMMARY} \label{summary}

In summary, the static and dynamic properties of the system for the EOF of an aqueous solution of NaCl between two similar parallel sheets made of either the black, blue, red, or green phosphorenes have been investigated by using the MD simulation. The EOF behavior has been studied by varying the surface charge density and the direction of the external electric field. In addition, the dependence of the thickness of the Stern layer on the surface charge density of the channel walls made by different allotropes has been considered. The results show that the thickness of the Stern layer increases with the increase in the absolute value of the surface charge density as well as the increase in the roughness ratio. Moreover, according to the MD simulations data, the Stern layer thickness is not significantly affected by the direction of the external electric field. 

Next, the effects of charge density, different allotropes as channel walls, and the direction of the external electric field on EOF velocity are examined. For all phosphorene allotropes and both electric field directions, the results indicate that the velocity increases with respect to the increase in the absolute value of the surface charge density in the DH region, and then decreases with increasing of $|\sigma|$ in the intermediate region. For high values of $|\sigma|$, the electroosmotic flow direction reverses with respect to the direction of the external electric field. 

For the electric field in the $y$--direction, due to the roughness effect, the mean value of the velocity is smaller than that of the $x$--direction. The EOF velocity in a channel with a blue phosphorene wall with the lowest roughness ratio is almost the same for both directions of the electric field. Finally, the reverse surface charge density $\sigma_r$ (the value of $\sigma$ at which the flow reverses) for all phosphorene allotropes and two directions of $x$ and $y$ for the external electric field is studied. The value of $|\sigma_r|$ increases as the surface roughness increases, except for the blue phosphorene, which has the least value of the surface roughness $R_{\textrm{s}}$ and is not sensitive to the direction of the external electric field. The values of $|\sigma_r|$ for the electric field in the $y$--direction are smaller than those in the $x$--direction, as the dynamics of the system are slower when the external electric field is applied in the $y$ direction compared to the case in which the external electric field is applied in the $x$ direction. This happens because the surface roughness on different phosphorene allotropes is uni-axial in the $x$ direction, i.e., the surface roughness is invariant in the $x$ direction.

The results of our study shed light on a better understanding of the dynamics of EOF inside the nanochannels made of different phosphorene allotropes.

%%===========================================================================================%%
%% If you are submitting to one of the Nature Portfolio journals, using the eJP submission   %%
%% system, please include the references within the manuscript file itself. You may do this  %%
%% by copying the reference list from your .bbl file, paste it into the main manuscript .tex %%
%% file, and delete the associated \verb+\bibliography+ commands.                            %%
%%===========================================================================================%%

%\bibliography{EOF-MoS2}% common bib file
%% if required, the content of .bbl file can be included here once bbl is generated
%%\input sn-article.bbl

%\scriptsize{
%\bibliography{rsc1} %You need to replace "rsc" on this line with the name of your .bib file
%\bibliographystyle{rsc} } %the RSC's .bst file

\end{document}